\documentclass[oribibl,envcountsame]{llncs}
\usepackage[T1]{fontenc}
\usepackage{ae}
\usepackage[english]{babel}
\usepackage[psamsfonts]{amsfonts}
\usepackage{amsfonts,amssymb,stmaryrd,amsmath}
\usepackage{url}
\usepackage{color} 

\newenvironment{mylist}%
        {\begin{list}{$\bullet$}%
	{\setlength{\leftmargin}{1cm}%
	 \setlength{\topsep}{0cm}}}%
        {\end{list}}

\newenvironment{mylist2}%
        {\begin{list}{$\centerdot$}%
	{\setlength{\leftmargin}{0.3cm}%
	 \setlength{\topsep}{0.1cm}}}%
        {\end{list}}

\newcounter{myitem}
\newenvironment{myenumerate}[1]%
        {\def\ender{#1}\begin{list}{\arabic{myitem}.}%
	{\usecounter{myitem}%
	 \setlength{\leftmargin}{0.7cm}%
	 \setlength{\labelsep}{0.2cm}%
	 \setlength{\labelwidth}{0.5cm}%
	 \setlength{\topsep}{0cm}}}%
        {\hfill\ender\medbreak\end{list}}

\newenvironment{mytheoreml}[1]%
      {\begin{theorem}{\bf #1}~\begin{myenumerate}{$\square$}}%
      {\end{myenumerate}\end{theorem}}

\newenvironment{mydefinitionl}[1]%
      {\begin{definition}{\bf #1}~\begin{myenumerate}{$\lozenge$}}%
      {\end{myenumerate}\end{definition}}

\newtheorem{hypothesis}{Hypotheses}
\newenvironment{myhypothesisl}[1]%
      {\begin{hypothesis}{\bf \hskip-.5em.~#1}\begin{myenumerate}{$\maltese$}}%
      {\end{myenumerate}\end{hypothesis}}

\newenvironment{mydefinition}[1]%
      {\begin{definition}{\bf #1}~}%
      {\hfill$\lozenge$\end{definition}}

\newenvironment{mytheorem}[1]%
      {\begin{theorem}{\bf #1}~}%
      {\hfill$\square$\end{theorem}}

      {\begin{remark}}%
      {\end{remark}}

\newcommand{\finthm}{\hfill$\square$\medbreak}
\newcommand{\finproof}{\hfill$\blacksquare$\medbreak}

\newcommand{\deltaeq}{\;\stackrel{\mbox{{\rm\tiny def}}}{=}\;}
\newcommand{\deltaiff}{\stackrel{\mbox{{\rm\tiny def}}}{\iff}}

\renewcommand{\vec}[1]{{\bf #1}}

\newcommand{\bigboxplus}{{\raisebox{-.25em}{\Large$\boxplus$}\,}}

\newcommand{\mydiv}{/}

\newcommand{\peq}{\:\dot{=}\:}
\newcommand{\pneq}{\:\dot{\neq}\:}
\newcommand{\psqcup}{\:\dot{\sqcup}\:}
\newcommand{\psqcap}{\:\dot{\sqcap}\:}
\newcommand{\pbot}{\dot{\bot}}
\newcommand{\widen}{\:\triangledown\:}
\newcommand{\pwiden}{\:\dot{\triangledown}\:}

\newcommand{\myi}{]\hskip-0.19em[}

\def\doframeit#1{\vbox{%
  \hrule height\fboxrule%
    \hbox{%
      \vrule width\fboxrule \kern\fboxsep%
      \vbox{\kern\fboxsep #1\kern\fboxsep }%
      \kern\fboxsep \vrule width\fboxrule }%
    \hrule height\fboxrule }}

\def\frameit{\smallskip \advance \linewidth by -7.5pt \setbox0=\vbox \bgroup%
\strut \ignorespaces }

\def\endframeit{\ifhmode \par \nointerlineskip \fi \egroup
\doframeit{\box0}}

\begin{document}
\sloppy
\renewcommand{\floatpagefraction}{0.8}
\setlength{\floatsep}{6pt}
\setlength{\textfloatsep}{10pt}


\title{A Few Graph-Based Relational\\ Numerical Abstract Domains\thanks{
This work was supported in part by the RTD project IST-1999-20527 "DAEDALUS" 
of the European IST FP5 program.}}

\author{Antoine Min\'e}
\institute{\'Ecole Normale Sup\'erieure de Paris, France,\\
           \email{mine@di.ens.fr},\\
           \url{http://www.di.ens.fr/~mine}}
\tocauthor{Antoine Min\'e (ENS Paris)}

\maketitle


\begin{abstract}
This article presents the systematic design of a class of relational
numerical abstract domains from non-relational ones.
Constructed domains represent sets of invariants of the form 
$(v_j-v_i\in C)$,
where $v_j$ and $v_i$ are two variables,
and $C$ lives in an abstraction of $\mathcal{P}(\mathbb{Z})$,
$\mathcal{P}(\mathbb{Q})$, or $\mathcal{P}(\mathbb{R})$.
We will call this family of domains {\em weakly relational domains}.
The underlying concept allowing this construction is an extension of
potential graphs and shortest-path closure algorithms in
exotic-like algebras.

Example constructions are given in order to retrieve well-known domains
as well as new ones.
Such domains can then be used in the Abstract Interpretation framework 
in order to design various static analyses.
A major benefit of this construction is its modularity, allowing to
quickly implement new abstract domains from existing ones.
\end{abstract}


\section{Introduction}

Proving the correctness of a program is essential, especially for
critical and embedded applications (such as planes, rockets, and so on).
Among several correctness criteria, one should ensure that a program can
never perform a run-time error (divide by zero, overflow, etc.).
A classical method consists in finding a {\em safety invariant}
before each dangerous operation in the program, and checking that the
invariant implies the good behavior of the subsequent operation.
Because this task is to be performed on the whole program---containing
maybe tens of thousands of lines---and must be repeated after even the
slightest code modification, we need a purely automatic
static analysis approach.

Discovering the tightest invariants of a program cannot be fully mechanized
in general, so we have to find some kind of sound approximation.
By sound, we mean that the analysis should find an over-approximation of the 
real invariant.
We will always discover {\em all} bugs in a program. 
However, we may find false alarms.

\section{Previous Work}

We will work in the well-known {\em Abstract Interpretation\/} framework,
proposed by Cousot and Cousot in \cite{ai,ai2}, which allows us to
easily describe sound and computable semantics approximations.

\subsection{Numerical Abstract Domains}

The crux of the method is to design a so-called {\em abstract domain}, that
is to say, a practical representation of the invariants we want
to study, together with a fixed set of operators and transfer
functions (union, intersection, widening, assignment, guard, etc.) used
to mimic the semantics of the programming language.

We will consider here {\em numerical abstract domains}.
Given the set $\mathcal{V}$ of the numerical variables of a program with
value in the set $\mathbb{I}$ 
(that can be $\mathbb{Z}$, $\mathbb{Q}$ or $\mathbb{R}$),
a numerical abstract domain will represent and manipulate subsets of
$\mathcal{V}\mapsto\mathbb{I}$.
Well-known {\em non-relational domains} include the {\em interval} domain
\cite{interv} (describing invariants of the form $v_i\in[c_1,c_2]$),
the {\em constant propagation} domain ($v_i=c$), and
the {\em congruence} domain \cite{cong} ($v_i\in a\mathbb{Z}+b$).
Well-known {\em relational domains}
include the {\em polyhedron} domain \cite{poly}
($\alpha_1v_1+\cdots+\alpha_nv_n\leq c$),
the {\em linear equality} domain \cite{karr}
($\alpha_1v_1+\cdots+\alpha_nv_n=c$), and
the {\em linear congruence equality} domain \cite{lincong}
($\alpha_1v_1+\cdots+\alpha_nv_n\equiv a\;[b]$).

Non-relational domains are fast, but suffer from poor precision:
they cannot encode relations between variables of a program.
Relational domains are much more precise, but also very costly.
Consider, for example, the simple program of Figure~\ref{examplefig} that
simulates many random walks and stores the hits in an array.
Our goal is to discover that, at program point $(\bullet)$, \textsf{x} 
is in the set $\{-5,-3,-1,1,3,5\}$ of allowed indices for \textsf{hit}, 
so that the instruction \textsf{hit[x]++} is correct.
The invariants found at $(\bullet)$ by several methods are shown in
Figure~\ref{resultfig}.
Remark that, even if the desired invariant is a simple 
combination of an interval and a congruence relation, 
all non-relational analyses fail
to discover it because they cannot infer the relationship between 
\textsf{x} and \textsf{i} at program point $(\bigstar)$.
It is often the case that, in order to find a given invariant at the end of
a loop, one must be able to express invariants of a more complex form inside
the loop.
In this example,
the desired result can be obtained by using relational analyses, as shown
in Figure~\ref{resultfig}.

\begin{figure}[tb]
\begin{frameit}\begin{center}
\vspace*{-0.5cm}
\begin{tabular}{ll}
\textsf{
\begin{tabular}{l}
hit: {\bf array} \{-5,-3,-1,1,3,5\} $\mapsto$ {\bf int};\\
{\bf for} k=1 {\bf to} 1000 {\bf do}\\
\quad  x=0;\\
\quad  {\bf for} i=1 {\bf to} 5 {\bf do}\\
\quad\quad    $(\bigstar)$ {\bf if} random() {\bf then} x++; {\bf else} x-\,-;\\
\quad {\bf done};\\
\quad  $(\bullet)$ hit[x]++;\\
{\bf done}\\
\end{tabular}
} 
&\qquad
\raisebox{-2.2cm}{\input{cfg.pstex_t}} 
\end{tabular}
\caption{A simple random walk program, and its control flow graph.}
\label{examplefig}
\end{center}\end{frameit}
\end{figure}

\begin{figure}[tb]
\begin{frameit}\begin{center}
\begin{tabular}{lc|c|c||c|c|}
\cline{3-6}
&& \;{\bf Interval}\; & \;{\bf Congruence}\; & {\bf Polyhedron} & \;{\bf Congruence equality}\\
\hline
\!\vline & $\;(\bigstar)\;$ & $i\in[1,5]$ & --- & $\;i\in[1,5]$, $-x\leq i-1\leq x\;$ & $x+i\equiv 1\;[2]$\\
\hline
\!\vline & $\;(\bullet)\;$ & --- & --- & $x\in[-5,5]$ & $x\equiv 1\;[2]$\\ 
\hline
\end{tabular}
\caption{Invariants discovered for the program in Figure~\ref{examplefig},
at program points $(\bullet)$ and $(\bigstar)$, using several non-relational 
(left) and relational (right) analyses.}
\label{resultfig}
\end{center}\end{frameit}
\end{figure}
\subsection{Graph-Based Algorithms}

Pratt remarked in \cite{pratt} that the satisfiability of a set of 
constraints of 
the form $(x-y\leq c)$ can be efficiently tested in $\mathbb{Z}$, 
$\mathbb{Q}$, or $\mathbb{R}$ by looking at the simple loops
of a directed weighted graph---so-called {\em potential graph\/}.
Shostak then extended in \cite{shostak} this graph-based algorithm to the 
satisfiability of constraints of the form $(\alpha x+\beta y\leq c)$, in
$\mathbb{Q}$ or $\mathbb{R}$.
Harvey and Stuckey proved in \cite{utvpi} that Shostak's algorithm can be
used to check satisfiability of constraints of the form 
$(\pm x\pm y\leq c)$ in $\mathbb{Z}$.
These approaches focus only on satisfiability and do not address the problem
of manipulating constraint sets.

Using Pratt's remark, the model-checking community developed a structure
called {\em Difference-Bound Matrix} (DBM, for short) and algorithms based
on {\em shortest-path closure} of weighted graphs to represent and 
manipulate constraints of the form $(x-y\leq c)$ and $(x\leq c)$.
DBMs are used to model-check timed-automata.
In \cite{periodic}, Toman and Chomicki introduced {\em periodicity graphs}
that manipulate constraints of the form $(x\equiv y+c\;[k])$, and apply this to
constraint logic programming and database query.

Unlike model-checking and constraint programming, we would like to
analyze generic programs, and not simply systems closed under restricted
constraint forms---such as timed automata, or database query languages.
Our methodology is first to choose an invariant form, and then to design
a fully-featured abstract domain (including guard and assignment transfer
functions, as well as a widening operator) allowing to discover invariants of 
this form on any program, using maybe coarse over-approximations for those
semantics functions that cannot be represented exactly using the chosen
invariant form.

In \cite{mine:padoII}, we already presented a DBM-based abstract domain 
allowing
to discover invariants of the form $(x-y\leq c)$ and $(x\leq c)$.
In \cite{mine:ast01}, we presented a slight extension, called the 
{\em octagon abstract domain}, allowing to discover invariants of the 
form $(\pm x\pm y \leq c)$.

\subsection{Our Contribution}

Our goal is to propose a new family of numerical abstract domains, based on
shortest-path closure algorithms, that allows to discover invariants
of the form $(x-y\in C)$, where $C$ lives in a non-relational domain.
This family generalizes the DBM-based abstract domain of \cite{mine:padoII} and
allows us to build new domains, such as the {\em zone congruence domain}
that discovers invariants of the form $(x\equiv y+c\;[k])$.
Such relational domains are between, in term of cost and
precision, non-relational domains and classical relational domains---such 
as polyhedron or congruence equality.
Thus, we will call these domains {\em weakly relational domains}.

We claim that such domains are useful as they give, on the example of 
Figure~\ref{examplefig}, almost the same result as the relational analyses, 
for a smaller cost.
Do not be fooled by the simplicity of this example program;
the abstract interpretation framework allows the design
of complex inter-procedural analyzes \cite{interproc} adapted to real-life
programming languages.
A numerical abstract domain is just a brick in the design
of an analysis; it can be plugged in many existing analyses, such as
pointer \cite{deutsch}, string cleanness \cite{dor}, 
termination analyses \cite{termin},
analyses of mobility \cite{feret}, probabilistic programs \cite{monniaux},
abstraction of tree-based semantics \cite{mauborgne}, etc.

\bigskip

The paper is organized as follows.
Section~3 reformulates the construction of non-relational numerical abstract
domains using the concept of {\em basis}.
Section~4 explains our generic construction of weakly relational domains
and applies it in order 
to retrieve the zone domain and build new abstract domains.
Section~5 provides a few applications and ideas for improvement.
We conclude in Section~6.
Important proofs are postponed to the Annex;
reading them may help to understand the definitions chosen in Sections~3-4 
(mainly Hypotheses~\ref{extbasisdef}).

\section{Bases and Non-Relational Numerical Abstract Domains}

This sections first recalls the concept of numerical abstract domain.
We introduce the new concept of {\em basis} and show how it can be used to
retrieve standard non-relational domains.
Introducing such a concept only for this purpose would be formalism for
the sake of formalism. However, we will show in the next section
how to use this concept to build our weakly relational domains.
Hence, bases are the common denominator between classical non-relational
domains and our weakly relational domain family.

From the implementation point of view, bases are modules sharing a common
signature, and we propose one functor for building non-relational
domains from this signature, and one functor for building weakly relational
domains.
From the mathematical point of view, this approach makes our proofs modular
and easier to handle.

\subsection{Semantics and Abstract Domains}

Let $P$ be a procedure-free, pointer-free program such as the one in 
Figure~\ref{examplefig}.
Let $\mathcal{V}=\{v_0,\ldots,v_{N-1}\}$ be
the set of its numerical variables, with values in the set
$\mathbb{I}$ (that can be $\mathbb{Z}$, $\mathbb{Q}$, or $\mathbb{R}$).
We attach to each node of the control flow graph of $P$ a set of
{\em environments}
${\bf e}^\natural\in\mathcal{D}^\natural\deltaeq\mathcal{P}(\mathbb{I}^N)$ 
that maps each variable to its value.
The information is propagated using the following equations:
\begin{mylist}
\item {\em guards}, corresponding to tests in the initial program, filter
the environment:
${\bf e}^\natural_{(\texttt{expr ?})}\deltaeq
\{\;(x_0,\ldots,x_{N-1})\in{\bf e}^\natural\;|\;
\texttt{expr}(x_0,\ldots,x_{N-1})\mbox{ holds}\;\}$;
\item {\em assignments} change the value of one variable:\\
${\bf e}^\natural_{(v_i\leftarrow \texttt{expr})}\deltaeq\{\;
(x_0,\ldots,\texttt{expr}(x_0,\ldots,x_{N-1}),\ldots)\;|\;(x_0,\ldots,x_i,\ldots)\in{\bf e}^\natural\;\}$;
\item {\em union} $\cup$ collects environments at control flow joins.
\end{mylist}

Because of loop constructs, the control flow graph contains loops and the
system of equations described above is recursive.
Classical safety semantics consider the {\em least fixpoint} solution.

\bigskip

This semantics is not decidable in general.
One thus constructs an {\em abstract domain} \cite{ai} which is a
computer-representable partially ordered set $(\mathcal{D},\preceq)$
connected to $(\mathcal{D}^\natural,\subseteq)$ by a monotonic
concretization function $\Gamma$.
Guard, assignment, and union operators have
{\em sound over-approximations} in $\mathcal{D}$, that is to say:
\begin{center}$
\left\{\begin{array}{lll}
\Gamma({\bf e}_{(\texttt{expr ?})}) & \supseteq\; & (\Gamma({\bf e}))^\natural_{(\texttt{expr ?})};\\
\Gamma({\bf e}_{(v_i\leftarrow\texttt{expr})}) & \supseteq\; & (\Gamma({\bf e}))^\natural_{(v_i\leftarrow\texttt{expr})};\\
\Gamma({\bf e}\sqcup {\bf f}) & \supseteq\; & \Gamma({\bf e})\cup\Gamma({\bf f})\enspace.
\end{array}\right.
$\end{center}

Unlike classical data-flow analysis, $\mathcal{D}$ can have an infinite
height, so one needs a {\em widening operator} \cite{ai} to compute,
in finite-time, an over-approximation of least fixpoints.
The widening operator $\widen$
should have the following properties \cite{ai}:
\begin{mydefinitionl}{Widening.}
\label{widendef}
\item $\forall x,y\in\mathcal{C},\;x\preceq x\widen y,\mbox{ and }
y\preceq x\widen y$.
\item For every increasing sequence $(y_n)_{n\in\mathbb{N}}$, 
the sequence $(x_n)_{n\in\mathbb{N}}$ defined by
$\left\{\begin{array}{lll}
x_0 & = & y_0,\\
x_{n+1} & = & x_n\widen y_n,
\end{array}\right.$
is ultimately stationary.
\hfill(Ascending chain condition.)\nobreak\par
\end{mydefinitionl}

The least fixpoint $\mbox{lfp}_{\bot} F$ of an abstract operator $F$
is replaced by the limit of the {\em stationary} sequence $X_0=\bot$, 
$X_{i+1}=X_i\widen F(X_i)$---see \cite{widen} for more information on when 
and how to use widenings.

It is a major result of abstract interpretation that, when computing in
the abstract domain with widenings, 
one obtains, in finite time, a sound over-approximation
of the initial semantics.

\subsection{Bases}
We call {\em basis} a structure that represents and manipulates
subsets of $\mathbb{I}$ in a way suitable to build a {\em non-relational}
abstract domain.
Such bases will be then used in the following section to build our family of
{\em relational} domains. It is given by:

\medbreak
\begin{mydefinitionl}{Basis.}
\label{basisdef}
\item A computer-representable set $\mathcal{C}$ with partial order
$\sqsubseteq$ and least element $\bot$.
\item A strict, monotonic, injective concretization 
$\gamma: \mathcal{C}\hookrightarrow\mathcal{P}(\mathbb{I})$.
\item Each element $C\subseteq\mathbb{I}$ has an over-approximation
$C^\sharp\in\mathcal{C}$: $\gamma(C^\sharp)\supseteq C$.
\item There exists an over-approximation $\sqcap$ for the intersection:
\begin{center}
$\gamma(C^\sharp_1\sqcap C^\sharp_2)\supseteq \gamma(C^\sharp_1)\cap \gamma(C^\sharp_2)\enspace.$
\end{center}
\item There exists an upper bound $\sqcup$:
\quad$C_1^\sharp,C_2^\sharp\sqsubseteq C_1^\sharp\sqcup C_2^\sharp$.
\item Each $k$-ary arithmetic expression 
$\texttt{expr}_k(c_1,\ldots,c_k)$ has an abstract over-approximated counterpart
$\texttt{expr}_k^\sharp(C^\sharp_1,\ldots,C^\sharp_k)$:
\begin{center}
$\gamma(\texttt{expr}_k^\sharp(C^\sharp_1,\ldots,C^\sharp_k))
\supseteq \{\;\texttt{expr}_k(c_1,\ldots,c_k)\;|\;
c_i\in\gamma(C^\sharp_i)\;\}\enspace.$
\end{center}
\item If $\mathcal{C}$ has strictly infinite chains, there is 
a {\em widening} operator $\widen$.
\end{mydefinitionl}

By strictness, $\gamma(\bot)=\emptyset$.
Thanks to points 2 and 3, there exists a unique abstract element
$\top$ such that $\gamma(\top)=\mathbb{I}$.
The least upper bound $\sqcup$ is also an over-approximation for the union:
$\gamma(C^\sharp_1\sqcup C^\sharp_2)\supseteq\gamma(C^\sharp_1)\cup\gamma(C^\sharp_2)$.

\subsection{A few Classical Bases}

\label{bases}

We now present a few set of bases that allow us to retrieve the
non-relational constant propagation \cite{ai}, interval \cite{interv}, and
congruence domains \cite{cong}.

\subsubsection{Constant Basis.}
$\mathcal{C}_{\mbox{cst}}\deltaeq
\{\bot,\;\top\}\cup\{\;c^\sharp\;|\;c\in\mathbb{I}\;\}.$

All abstract operators are straightforward and not discussed here
(see \cite{ai} for more details).
There is no need for a widening operator.

\subsubsection{Interval Basis.}
$\mathcal{C}_{[a,b]}\deltaeq\{\bot\}\cup\{\;[a,b]\;|\;a\in\mathbb{I}\cup\{-\infty\},\;b\in\mathbb{I}\cup\{+\infty\},\;a\leq b\footnote{Bounds are part of the interval only if finite. Do not be confused by closed interval notations such as $[a,+\infty]$; the interval cannot contain infinite elements.}\;\}.$

Most abstract operators are straightforward (see \cite{interv}
for more details).
We will only recall here the classical widening operator:
\begin{center}$
[a_1,b_1]\widen[a_2,b_2]\deltaeq
\left[
\left\{\begin{array}{ll}
a_1 & \mbox{ if }a_1\leq a_2\\
-\infty\; & \mbox{elsewhere}
\end{array}\right.,\quad
\left\{\begin{array}{ll}
b_1 & \mbox{ if }b_1\geq b_2\\
+\infty\; & \mbox{elsewhere}
\end{array}\right.
\right]\enspace.
$\end{center}

In $\mathbb{Q}$ or $\mathbb{R}$, one can alternatively
define the {\em open interval lattice} $\mathcal{C}_{]a,b[}$ the same way.
One can even combine these informations to obtain a basis
$\mathcal{C}_{\myi a,b\myi}$ where each bound may or may not be included.

\subsubsection{Congruence Basis.}
$\mathcal{C}^{\mathbb{Z}}_{a\mathbb{Z}+b}\deltaeq
\{\bot\}\cup\{\;(a\mathbb{Z}+b)\;|\;a\in\mathbb{N}^*\cup\{\infty\},\;b\in\mathbb{Z}\;\}.$

A basis is built on $\mathcal{C}^{\mathbb{Z}}_{a\mathbb{Z}+b}$ thanks to the
operators described in Figure~\ref{congruencefig2} (using the definitions
of Figure~\ref{congruencefig1}).
However, for the sake of conciseness, Figure~\ref{congruencefig2} does not
present abstract $k$-ary arithmetic expressions,
but the binary plus, which
is denoted by the infix $\boxplus$ operator (see \cite{cong} for more details).
There is no strictly increasing infinite chain, so there is no need for
a widening operator.

One may also consider to adapt the definitions of Figures~\ref{congruencefig1}
and \ref{congruencefig2} to {\em rational congruences} 
\cite{ratcong}:
$\mathcal{C}^{\mathbb{Q}}_{a\mathbb{Z}+b}=
\{\bot\}\cup\{\;(a\mathbb{Z}+b)\;|\;a\in\mathbb{Q}^{>0}\cup\{\infty\},\;b\in\mathbb{Q}\;\}.$

\begin{figure}[tb]
\begin{frameit}
\smallskip\par
\qquad In the following, $x,x'\in\mathbb{Z}$
and $y,y'\in\mathbb{N}^*\cup\{\infty\}$:
\begin{mylist}
\item $y \mydiv y' \deltaiff y$ is a divisor of $y'$
($\exists k\in\mathbb{N}^*$ such that $y'=ky$), or $y'=\infty$;
\item $x\equiv x'\;[y]\deltaiff x\neq x'\mbox{ and }y\mydiv |x-x'|,\mbox{ or }x=x'$;
\item $\vee$ is the least common multiple, extended by
$y\vee\infty\deltaeq\infty\vee y\deltaeq\infty$;
\item $\wedge$ is the greatest common divisor, extended by
$y\wedge\infty\deltaeq\infty\wedge y\deltaeq y$.
\end{mylist}
\caption{Classical arithmetic operators extended to $\mathbb{N}^*\cup\{\infty\}$.}
\label{congruencefig1}
\end{frameit}
\end{figure}

\begin{figure}[tb]
\begin{frameit}\begin{center}
\begin{mylist}
\item {\em Concretization:}

\qquad
$\gamma(C)\deltaeq\left\{\begin{array}{ll}
\{\;ak+b\;|\;k\in\mathbb{Z}\;\}\quad &
\mbox{if }C=(a\mathbb{Z}+b),\;a\neq\infty;\\
\{\;b\;\} & \mbox{if }C=(\infty\mathbb{Z}+b);\\
\emptyset & \mbox{if }C=\bot.
\end{array}\right.$
\item 
{\em Order:}
\begin{mylist2}
\item 
$(a\mathbb{Z}+b)\sqsubseteq(a'\mathbb{Z}+b')\deltaiff
a'\mydiv a\mbox{ and }b\equiv b'\;[a']$.
\item 
$\bot\sqsubseteq C,\;\forall C\in\mathcal{C}$.
\end{mylist2}
\item
{\em Intersection (exact abstract counterpart for the intersection $\cap$):}
\begin{mylist2}
\item 
$(a\mathbb{Z}+b)\sqcap(a'\mathbb{Z}+b')\deltaeq
\left\{\begin{array}{ll}
(a\vee a')\;\mathbb{Z}+b''\quad & \mbox{if }b\equiv b'\;[a\wedge a'],\\
\bot & \mbox{elsewhere},\\
\end{array}\right.$\\
\qquad where $b''$ is such that $b''\equiv b\;[a\vee a']\equiv b'\;[a\vee a']$
(Bezout Theorem).
\item 
$\bot\sqcap C\;\deltaeq\;C\sqcap\bot\;\deltaeq\;\bot,\;\forall C\in\mathcal{C}$.
\end{mylist2}
\item 
{\em Least Upper Bound:}
\begin{mylist2}
\item
$(a\mathbb{Z}+b)\sqcup(a'\mathbb{Z}+b')\;\deltaeq\;
(a\wedge a'\wedge |b-b'|)\;\mathbb{Z}+\min(b,b')$.
\item
$\bot\sqcup C\;\deltaeq\;C\sqcup\bot\;\deltaeq\;C,\;\forall C\in\mathcal{C}$.
\end{mylist2}
\item 
{\em Sum (exact abstract counterpart for the binary + operator):}
\begin{mylist2}
\item
$(a\mathbb{Z}+b)\boxplus(a'\mathbb{Z}+b')\;\deltaeq\;
(a\wedge a')\;\mathbb{Z}+(b+b')$.
\item
$\bot\boxplus C\;\deltaeq\;C\boxplus\bot\;\deltaeq\;\bot,\;\forall C\in\mathcal{C}$.
\end{mylist2}
\end{mylist}
\caption{Concretization and abstract operators in $\mathcal{C}^{\mathbb{Z}}_{a\mathbb{Z}+b}$.}
\label{congruencefig2}
\end{center}\end{frameit}
\end{figure}

\subsection{Building Non-relational Domains from Bases}

Building a non-relational domain $(\mathcal{D},\preceq)$ from a basis 
$(\mathcal{C},\sqsubseteq)$ is straightforward:
\begin{mylist}
\item We set $\mathcal{D}\deltaeq\mathcal{V}\mapsto\mathcal{C}$.
\item The concretization $\Gamma$, order $\preceq$, union $\sqcup$, and 
widening $\widen$ are simply
point-wise versions of the corresponding operators on the basis.
\item Assignments are defined using the abstract counterpart of expressions:
\begin{center}$
(C_1^\sharp,\ldots,C_i^\sharp,\ldots)_{(v_i\leftarrow\texttt{\em expr})}
\deltaeq(C_1^\sharp,\ldots,\texttt{\em expr}^\sharp_{N}(
C_1^\sharp,\ldots,C_{N-1}^\sharp),\ldots)
\enspace.
$\end{center}
\item Only non-relational guards $(v_i\in C\;?)$ do some filter job:
\begin{center}$
(C_1^\sharp,\ldots,C_i^\sharp,\ldots)_{(v_i\in C\;?)}\deltaeq
(C_1^\sharp,\ldots,C_i^\sharp\sqcap C^\sharp,\ldots)
\mbox{ where }\gamma(C^\sharp)\supseteq C\enspace.
$\end{center}
In other guard cases, it is safe to use the identity function.
\end{mylist}

\bigskip

From an implementation point of view, the non-relational domain
is simply a generic functor module, and each basis implementation
is a module.

\section{Building Weakly Relational Domains from Bases}

Now we would like to represent relations of the form
$v_j-v_i\in \gamma(C)$ where $C$ lives in a basis $\mathcal{C}$
(instead of $v_i\in \gamma(C)$).
A plain basis is not sufficient, we will need a 
way---a so-called {\em closure}---to propagate relational information.
The main result of this paper can be schemed as follows:
\begin{center}
\fbox{\begin{tabular}{ccccc}
\begin{tabular}{c}basis\\(with extra hypotheses)\end{tabular}
& \quad + \quad & closure & \quad $\Longrightarrow$  \quad & 
\begin{tabular}{c}weakly relational\\ domain\end{tabular}
\end{tabular}}
\end{center}

\subsection{Hypotheses on the Basis}
Not all bases $\mathcal{C}$ are acceptable.
We need the following extra hypotheses:
\begin{myhypothesisl}{Acceptable Bases.}
\label{extbasisdef}
\item There exists {\em exact} abstract counterparts for the intersection 
$\sqcap$ (which should also be a {\em lower bound} for $\sqsubseteq$), 
unary minus $\boxminus$, and binary plus $\boxplus$ operators:
\begin{mylist2}
\item 
$\gamma(x\boxplus y)=\{\;a+b\;|\;a\in\gamma(x),\;b\in\gamma(y)\;\}$;
\hfill(Abstract plus.)
\item
$\gamma(\boxminus x)=\{\;-a\;|\;a\in\gamma(x)\;\}$;
\hfill(Abstract opposite.)
\item
$x\sqcap y\sqsubseteq x,y$, so
$\gamma(x\sqcap y)=\gamma(x)\cap\gamma(y)$.
\hfill(Abstract intersection.)
\end{mylist2}
\item Each singleton $\{c\},\;c\in\mathbb{I}$ must be 
{\em exactly} represented by an abstract element $c^\sharp\in\mathcal{C}$:
$\gamma(c^\sharp)=\{c\}$.
\item For each finite family $(x_i)_{i\in I}$,
$$\bigsqcap_{i\in I} x_i=\bot\Longrightarrow
\exists i,j\in I,\;x_i\sqcap x_j=\bot\enspace.$$
\item $\sqcap$ distributes $\boxplus$: for each family $(x_i)_{i\in I}$ and
element $x$ of $\mathcal{C}$,
$$\mbox{if }\bigsqcap_{i\in I}x_i\neq\bot,\mbox{ then }
\bigsqcap_{i\in I} (x\boxplus x_i)=
x\boxplus\left(\bigsqcap_{i\in I}x_i\right)\enspace.$$
\item $\boxminus$ distributes $\boxplus$ and $\sqcap$.
$\boxplus$ and $\sqcap$
are commutative and associative.
\end{myhypothesisl}

These hypotheses were stated in order to prove our main theorem, which is
the correctness of the closure operator.
Thus, one may have to wait until Theorem~\ref{closurethm}--- and its proof 
postponed in Annex A---
Remark that Hypotheses~\ref{extbasisdef}.3-4 are very strong.
The full basis $\mathcal{C}=\mathcal{P}(\mathbb{I})$, for instance, does
not respect them.

\subsubsection{Remark.} 
There exists resemblance between bases respecting Hypotheses~\ref{extbasisdef}
and the graph-theory classical notion of 
{\em complete d\"ioid} \cite{dioid}---an extension of {\em exotic algebras}.
A complete d\"ioid is a complete semi-lattice with an 
addition (our $\sqcap$), and a multiplication (our $\boxplus$) that
distributes over the addition.
However, full distributivity in d\"ioids implies that 
$\bot\boxplus\top=\top$
where we would have preferred $\bot\boxplus\top=\bot$.
Thus, in our framework, distributivity is restricted 
(Hypothesis~\ref{extbasisdef}.4).

\subsection{Representing Relations}
A set of constraints of the form $v_j-v_i\in \gamma(C),\;C\in\mathcal{C}$
is now represented by a {\em coherent constraint matrix\/}:
\begin{mydefinitionl}{Constraint Matrices.}
\item
A {\em constraint matrix} $\vec{m}$ is a $N\times N$ matrix with elements in 
$\mathcal{C}$; 
the element $\vec{m}_{ij}$ represents the constraint 
$v_j-v_i\in\gamma(\vec{m}_{ij})$.
\item We suppose, as an implicit constraint, that $v_0=0$, so that
unary constraints $v_i\in\gamma(\vec{m}_{0i})$ can be represented as
$v_i-v_0\in\gamma(\vec{m}_{0i})$.
\item $\vec{m}$ is {\em coherent} if 
$\forall i,j,\;\vec{m}_{ij}=\boxminus\:\vec{m}_{ji}$ and
$\forall i,\;\gamma(\vec{m}_{ii})=\{0\}$.
\item $\vec{m}$ represents the set 
(so-called {\em concretization} of $\vec{m}$):\\
$\Gamma(\vec{m})
\deltaeq\{\;(x_0,\ldots,x_{N-1})\in\mathbb{I}^N\;|\;
x_0=0,\;\forall i,j,\;x_j-x_i\in\gamma(\vec{m}_{ij})\;\}\enspace.$
\end{mydefinitionl}

Our abstract domain is the set $\mathcal{D}$ of coherent constraint 
matrices, ordered by the point-wise extension $\preceq$ of the 
partial order $\sqsubseteq$ on $\mathcal{C}$:
\begin{center}$\begin{array}{cc}
\vec{m}\preceq\vec{n}\deltaiff 
\forall i,j,\;\vec{m}_{ij}\sqsubseteq\vec{n}_{ij};\\
\vec{m}\peq\vec{n}\deltaiff 
\forall i,j,\;\vec{m}_{ij}=\vec{n}_{ij};\\
\pbot\deltaeq\inf_{\preceq}\mathcal{D}\mbox{ is such that }
\forall i,j,\;\pbot_{ij}=\bot\enspace.
\end{array}$\end{center}

The concretization function on $\mathcal{D}$ is $\Gamma$, and we have:
\begin{mytheoreml}{Monotony of $\Gamma$.}
\item $\vec{m}\preceq\vec{n}\Longrightarrow
\Gamma(\vec{m})\subseteq\Gamma(\vec{n})$.
\item $\vec{m}\peq\vec{n}\Longrightarrow
\Gamma(\vec{m})=\Gamma(\vec{n})$.
\end{mytheoreml}

However, this is {\em not an equivalence} and we can have two different 
constraint matrices $\vec{m}\pneq\vec{n}$ with the same concretization
$\Gamma(\vec{m})=\Gamma(\vec{n})$.

\subsection{General Closure Operator}

\subsubsection{Implicit Constraints.}
Because our abstract domain is relational, the constraints between variables 
are not independent.
One can deduce a constraint on $x-z$ by adding a constraint on 
$x-y$ to a constraint on $y-z$.
Such deduced constraints are called {\em implicit constraints} because
they may not be present explicitly in $\vec{m}$.
More generally, given any 
path $\langle i=i_1,\ldots,i_n=j\rangle$ in $\vec{m}$, 
we can construct the following implicit constraint:
\begin{center}
$v_j-v_i\in\gamma\left(\bigboxplus_{l=1}^{n-1} \vec{m}_{i_l\:i_{l+1}}\right)
\enspace.$
\end{center}

\subsubsection{Shortest-Path Closure.}
A nice property of DBMs \cite{dbm} and periodicity graphs \cite{periodic} that
will hold for our constraint matrices is that the concretization is entirely
determined by the set of implicit constraints of the above form.
DBMs use any {\em shortest-path closure} algorithm in order to make all
implicit constraints explicit.
Here, we adapt the {\em Floyd-Warshall algorithm}
\cite[\S 25.3]{CLR}, to our matrices.

\begin{mydefinition}{Closure.}
Let $\vec{m}$ be a coherent matrix.
Its {\em closure} is the result $\vec{m}^{\bigstar}$ of the following 
{\em modified Floyd-Warshall algorithm}:
\begin{center}$\left\{\begin{array}{lll}
\vec{m}^0 & \deltaeq & \vec{m};\\
\vec{m}^{k+1}_{ij} & \deltaeq & \vec{m}^k_{ij}\sqcap(\vec{m}^k_{ik}\boxplus\vec{m}^k_{kj});\\
\vec{m}^{\bigstar} & \deltaeq & \vec{m}^N\enspace.
\end{array}\right.$\end{center}
\end{mydefinition}

The Floyd-Warshall algorithm was chosen because it is easy to understand,
straightforward to implement, and easy to adapt to constraint matrices.
It performs $\mathcal{O}(N^3)$ elementary basis operations.

Here is the main theorem of this paper.
The following results will be
used extensively in Section~\ref{opsect} in order to design 
abstract operators.
The proof of this theorem relies heavily on Hypotheses~\ref{extbasisdef}---in 
fact, the proof itself motivated the hypotheses.

\begin{mytheoreml}{Closure.}
\label{closurethm}
\item $\Gamma(\vec{m}^{\bigstar})=\Gamma(\vec{m})$.
\item $\Gamma(\vec{m})=\emptyset\iff\exists i,\;\vec{m}^{\bigstar}_{ii}=\bot$.
\item If $\Gamma(\vec{m})\neq\emptyset$, then $\vec{m}^{\bigstar}$ enjoys the 
following properties:
\begin{mylist2}
\item $\vec{m}^\bigstar$ is a coherent matrix;
\hfill(Coherence.)
\item $\forall i,j,\;\vec{m}^{\bigstar}_{ij}=
\bigsqcap_{\langle i=i_1,\ldots,i_n=j\rangle} 
\bigboxplus_{l=1}^{n-1}\vec{m}_{i_l\:i_{l+1}}$;
\hfill(Transitive closure.)
\item
$\forall i,j,\;\forall c\in\gamma(\vec{m}^{\bigstar}_{ij}),\;
\exists(x_0,\ldots,x_{N-1})\in\Gamma(\vec{m}),\;x_j-x_i=c$;
\hfill(Saturation.)
\item 
$\vec{m}^{\bigstar}=
\inf_{\preceq}\{\;\vec{n}\;|\;\Gamma(\vec{m})=\Gamma(\vec{n})\;\}$;
\hfill(Normal form.)
\item $\vec{m}^{\bigstar\bigstar}=\vec{m}^\bigstar$.
\hfill(Closure.)
\end{mylist2}
\end{mytheoreml}

\subsubsection{Incremental Closure.}
When modifying slightly a closed matrix, we do not need to perform the 
modified Floyd-Warshall algorithm completely to get the closure of the
new matrix.
If the upper-left $M\times M$ sub-matrix of $\vec{m}$ is already closed, we can
use the following $\mathcal{O}((N-M)\cdot N^2)$ algorithm:
\begin{center}$\left\{\begin{array}{llll}
\vec{m}^0 & \deltaeq & \vec{m};\\
\vec{m}^{k+1}_{ij} & \deltaeq & \vec{m}^k_{ij}&\mbox{ if }i,j,k<M;\\
\vec{m}^{k+1}_{ij} & \deltaeq & \vec{m}^k_{ij}\sqcap(\vec{m}^k_{ik}\boxplus\vec{m}^k_{kj})\quad&\mbox{ elsewhere};\\
\vec{m}^{\bigstar} & \deltaeq & \vec{m}^N\enspace.
\end{array}\right.$\end{center}
We can adapt easily the algorithm---permuting variables---to get a general
incremental closure algorithm performing
$\mathcal{O}(N^2\cdot c)$ elementary basis operations, where
$c$ is the number of lines and columns that have changed since the last 
closure.

\subsection{Generic Operators}
\label{opsect}

\subsubsection{Emptiness Testing.}
Testing the satisfiability of a constraint matrix is done using 
Theorem~\ref{closurethm}.2.
Unlike the constraint programming approach, we do not use a specific loop-based
satisfiability algorithm, but let our generic closure algorithm solve both
the satisfiability and the normal form problems at once.

\subsubsection{Equality, Inclusion Testing.}
The normal form property of Theorem~\ref{closurethm}.3 allows us to
easily test equality and inclusion of non-empty concretizations:
\begin{mytheoreml}{Equality and Inclusion Testing.}
\item $\Gamma(\vec{m})=\Gamma(\vec{n})\iff
\vec{m}^{\bigstar}\peq\vec{n}^{\bigstar}$.
\item $\Gamma(\vec{m})\subseteq\Gamma(\vec{n})
\iff \vec{m}^{\bigstar}\preceq\vec{n}$.
\end{mytheoreml}
Remark that we do not need to close the right argument while testing
inclusion.

\subsubsection{Union, Intersection.}
$\gamma(\mathcal{C})$ is stable under intersection, so we simply extend 
point-wisely
$\sqcap$ to represent the intersection of two concretizations:
\begin{center}
$\left[\vec{m}\psqcap\vec{n}\right]_{ij}\deltaeq\vec{m}_{ij}\sqcap\vec{n}_{ij}
\enspace.$\end{center}
\begin{mytheorem}{Intersection.}
$\Gamma(\vec{m}\psqcap\vec{n})=\Gamma(\vec{m})\cap\Gamma(\vec{n})\enspace.$
\end{mytheorem}
$\gamma(\mathcal{C})$ is not generally closed under union,
neither is $\Gamma(\mathcal{D})$.
However, if there exists an upper bound $\sqcup$ in $\mathcal{C}$, we can 
extend it point-wisely in $\mathcal{D}$:
\begin{center}
$\left[\vec{m}\psqcup\vec{n}\right]_{ij}\deltaeq\vec{m}_{ij}\sqcup\vec{n}_{ij}
\enspace.$\end{center}
If $\sqcup$ is a {\em least} upper bound, $\psqcup$ can be used to determine 
the least upper bound of two concretizations, 
{\em provided the arguments are closed matrices}.
\begin{mytheoreml}{(Least) Upper Bound.}
\item If $\forall a,b\in\mathcal{C},\;\gamma(a\sqcup b)\supseteq\gamma(a)\cup
\gamma(b)$,\\then
$\Gamma(\vec{m}\psqcup\vec{n})\supseteq\Gamma(\vec{m})\cup\Gamma(\vec{n}).$
\hfill(Upper bound.)
\item If $\gamma(a\sqcup b)=\inf_{\subseteq}
\{\;\gamma(c)\;|\;\gamma(c)\supseteq\gamma(a)\cup\gamma(b)\;\}$, then\\
$\Gamma(\vec{m}^{\bigstar}\psqcup \vec{n}^{\bigstar})=\inf_{\subseteq}
\{\;\Gamma(\vec{o)}\;|\;\Gamma(\vec{o)}\supseteq\Gamma(\vec{m)}\cup\Gamma(
\vec{n)}\;\}.$
\hfill(Least upper bound.)
\item $(\vec{m}^{\bigstar}\psqcup\vec{n}^{\bigstar})^{\bigstar}=
\vec{m}^{\bigstar}\psqcup\vec{n}^{\bigstar}.$
\hfill($\psqcup$ respects closure.)\nobreak\par
\end{mytheoreml}

\subsubsection{Widening.}
$\mathcal{D}$ has infinite strictly increasing chains only if 
$\mathcal{C}$ has.
A widening $\pwiden$ 
is obtained on $\mathcal{D}$ by point-wise application
of the widening $\widen$ on $\mathcal{C}$:
\begin{center}
$\left[\vec{m}\pwiden\vec{n}\right]_{ij}\deltaeq\vec{m}_{ij}\widen\vec{n}_{ij}
\enspace.$\end{center}
$\pwiden$ respects Definition~\ref{widendef}.
Thus, the least fixpoint of an operator $F$ can be over-approximated by the 
limit of the stationary sequence $X_{i+1}=X_i\pwiden F(X_i)$.
One could expect, as for the least upper bound, to get a better precision by
closing the arguments of $\pwiden$, but this is not the case.
Even worse, enforcing the closure of the chain by computing
$X_{i+1}=(X_i\pwiden F(X_i))^\bigstar$ 
{\em breaks the ascending chain condition} and prevents the analysis from 
terminating in some cases.
We advocate here the use of the following iteration:
$X_{i+1}=X_i\pwiden F(X_i^\bigstar)$.

\subsubsection{Guard.}
We can easily implement tests of the form $(v_j-v_i\in C\;?)$:
\begin{center}$
\begin{array}{l}
\left[\vec{m}_{(v_{j}-v_{i}\in C\;?)}\right]_{kl}\deltaeq
\left\{\begin{array}{ll}
\vec{m}_{kl}\sqcap C^\sharp & \quad\mbox{if }(k,l)=(i,j);\\
\vec{m}_{kl}\sqcap (\boxminus C^\sharp) & \quad\mbox{if }(k,l)=(j,i);\\
\vec{m}_{kl} & \quad\mbox{elsewhere};
\end{array}\right.\\
\mbox{choosing }C^\sharp\mbox{ such that }\gamma(C^\sharp)\supseteq C\enspace.
\end{array}
$\end{center}
Tests of the form $(v_j\in C\;?)$ are implemented by choosing $i=0$.\\
For other tests, it is safe to do nothing:
\begin{center}$
\vec{m}_{(?)}\deltaeq\vec{m}\enspace.
$\end{center}

\subsubsection{Projection.}
In order to find the set of values that a variable can take, we use the
following theorem derived from the saturation property of the closure:
\begin{mytheorem}{}
\label{projthm}
$\{\;x\;|\;\exists (x_0,\ldots,x_{N-1})\in\Gamma(\vec{m})\mbox{ with }
x_i=x\;\}
=\gamma(\vec{m}^\bigstar_{0i})$\enspace.
\end{mytheorem}

\subsubsection{Forget.}
Forgetting the value of a variable is useful to implement the random
assignment $(v_i\leftarrow ?)$, which also serves as a coarse approximation 
for complex assignments.
Before forgetting all information on a variable, one should
close the argument matrix so that we do not loose implicit constraints:
\begin{center}$
\left[\vec{m}_{(v_{i}\leftarrow ?)}\right]_{kl}\deltaeq
\left\{\begin{array}{ll}
\top & \quad\mbox{if }k=i\mbox{ or }l=i;\\
\vec{m}^{\bigstar}_{kl}
& \quad\mbox{elsewhere}\enspace.
\end{array}\right.
$\end{center}
\begin{mytheorem}{}
\label{forgetthm}\\
$\Gamma(\vec{m}_{(v_i\leftarrow ?)})=\{\;(x_0,\ldots,x_i,\ldots)\;|\;
\exists x,\;
(x_0,\ldots,x,\ldots)\in\Gamma(\vec{m})\;\}$\enspace.
\end{mytheorem}

\subsubsection{Assignment.}
For assignments of the form $(v_i\leftarrow v_j+c)$, one can find an 
{\em exact}
abstract counterpart:
\begin{center}$ 
\begin{array}{l}
\left[\vec{m}_{(v_i\leftarrow v_i+c)})\right]_{kl}\deltaeq
\left\{\begin{array}{ll}
\vec{m}_{kl}\boxplus \{c\}&\mbox{if }k=i\mbox{ and }l\neq i;\\
\vec{m}_{kl}\boxplus \{-c\}\quad&\mbox{if }l=i\mbox{ and }k\neq i;\\
\vec{m}_{kl}&\mbox{elsewhere};\\
\end{array}\right.
\\ \\
\vec{m}_{(v_i\leftarrow v_j+c)}\deltaeq
(\vec{m}_{(v_i\leftarrow ?)})_{(v_i-v_j\in\{c\}
\;?)}\qquad\mbox{ when }i\neq j\enspace.
\end{array}
$\end{center}

For generic assignments $(v_i\leftarrow \texttt{expr}(v_1,\ldots,v_{N-1}))$, 
one can always fall back to imprecise
non-relational analysis, first projecting the
variables, then using the abstraction $\texttt{expr}^\sharp$ of \texttt{expr}
in our basis:
\begin{center}$
\begin{array}{l}
\vec{m}_{(v_i\leftarrow \texttt{expr}(v_1,\ldots,v_{N-1}))}\deltaeq
(\vec{m}_{(v_i\leftarrow ?)})_{(v_i\in \gamma(C^\sharp)\;?)}\\
\mbox{where } 
C^\sharp\deltaeq
\texttt{expr}^\sharp(\vec{m}^\bigstar_{01},\ldots,\vec{m}^\bigstar_{0(N-1)})
\enspace.
\end{array}
$\end{center}

Trying to be the most precise in all cases may lead to complex algorithms.
It seems only worth trying to be a little more precise in some widespread 
cases, such as $(v_i\leftarrow v_j+v_k)$, for instance:
\begin{center}$
\vec{m}_{(v_i\leftarrow v_j+v_k)}\deltaeq
(\vec{m}_{(v_i\leftarrow\;?)})_
{(v_i\in\gamma(\vec{m}^\bigstar_{0j}\boxplus\vec{m}^\bigstar_{0k})\;?)\;
 (v_i-v_j\in\gamma(\vec{m}^\bigstar_{0k})\;?)\;
 (v_i-v_k\in\gamma(\vec{m}^\bigstar_{0j})\;?)}\enspace.
$\end{center}

\subsubsection{Interaction with the Closure.}
Some of the above operators require the matrix argument(s) to be closed.
Some do respect closure---the result is closed if the argument(s) 
is(are)---and some do not (intersection, guard, assignment, etc.).
We thus advocate the use of a {\em lazy} method that remembers when a matrix 
is in closed form, and recomputes the closure only when needed.
When only a few lines and columns of the matrix are changed 
(guard, assignment, etc.), we can
use the incremental closure. It is useless
when all coefficients are changed at once (intersection, widening).

\subsection{Some Constructed Domains}

We are now ready to apply our construction to the bases presented in 
Section~\ref{bases}, thanks to the following theorem:
\begin{mytheorem}{}
$\mathcal{C}_{\mbox{\em cst}}$, $\mathcal{C}_{[a,b]}$,
$\mathcal{C}_{\myi a,b\myi}$, $\mathcal{C}^{\mathbb{Z}}_{a\mathbb{Z}+b}$, and
$\mathcal{C}^{\mathbb{Q}}_{a\mathbb{Z}+b}$ respect 
Hypotheses~\ref{extbasisdef}.
\end{mytheorem}

\subsubsection{Translated Equality Domain.}
The simplest domain is obtained from the constant basis 
$\mathcal{C}_{\mbox{cst}}$
and represents constraints of the form $(v_i=v_j+c)$.
This domain is not of great practical interest:
its expressive power is low as it is a particular case of
the following two domains.
It is possible that more efficient solutions exist, as we are not very far
from simple equality constraints $v_i=v_j$ for which
very efficient algorithms are known (such as, the {\em Union-Find}
algorithm \cite[\S 22]{CLR}).

\subsubsection{Zone Domain.}
In order to represent invariants of the form $(v_i-v_j\leq c)$, one can think
of the basis of initial segments 
$\{\;]-\infty,a]\;|\;a\in\mathbb{I}\cup\{+\infty\}\}$,
but initial segments are not closed under the $\boxminus$ operation
(Hypothesis~\ref{extbasisdef}.1).
Completing this basis, one naturally find the interval basis
$\mathcal{C}_{[a,b]}$.

Compared to classical DBMs \cite{mine:padoII}, the domain obtained is a little
redundant (each constraint is represented twice), but has exactly the same
expressiveness and complexity.
It has the advantage of being implemented over any existing interval
library, greatly reducing the need for programming.
One can also enhance the zone domain in $\mathbb{Q}$ and $\mathbb{R}$ 
using the $\mathcal{C}_{\myi a,b\myi}$ basis that manipulates both strict and
non strict constraints.

\subsubsection{Zone-Congruence Domain.}
Using the integer congruence basis $\mathcal{C}^{\mathbb{Z}}_{a\mathbb{Z}+b}$, one builds a domain that discovers constraints of the form
$(v_i-v_j\equiv a\;[b])$.
This construction looks like {\em periodicity graphs} \cite{periodic},
but we treat here the case of least upper bound and general purpose
transfer functions in detail, whereas \cite{periodic} is only interested in
satisfiability, normal form and conjunction.
Moreover, we feel that \cite{periodic} misses the correct proof of 
the normal form theorem (our Theorem~\ref{closurethm}) and does not understand
that it relies on some strong properties of congruence sets 
(Hypotheses~\ref{extbasisdef}).
Our framework can also extend this domain to a domain of 
{\em rational congruences}: $(v_i-v_j\equiv a\;[b])$ with $a,b\in\mathbb{Q}$.

\subsubsection{Product Domain.}
{\em Reduced product} is a well-known technique \cite{ai}
for improving the precision of an analysis by combining the power of
two abstract domains.
It often gives better results than two
separate analyses, because it conveys information from one domain to
the other during the analysis via a so-called 
{\em reduction procedure}, which is a couple of binary operators
$(\circlearrowleft,\circlearrowright)$ such that:
\begin{center}$
\begin{array}{cc}
\begin{array}{l}
\circlearrowleft: \mathcal{D}_1\times\mathcal{D}_2\mapsto\mathcal{D}_1;\\
\circlearrowright: \mathcal{D}_1\times\mathcal{D}_2\mapsto\mathcal{D}_2;\\
\end{array}
&\quad
\left\{\begin{array}{l}
C_1\circlearrowleft C_2\preceq_1 C_1;\\
C_1\circlearrowright C_2\preceq_2 C_2;\\
\Gamma_1(C_1\circlearrowleft C_2)\cap\Gamma_2(C_1\circlearrowright C_2)=
\Gamma_1(C_1)\cap\Gamma_2(C_2)\enspace.
\end{array}\right.
\end{array}
$\end{center}

In our case, the reduction can be defined on bases---as 
long as Hypotheses~\ref{extbasisdef} are not broken---with the exact same
precision benefit.
Moreover, reductions are easier to design on non-relational bases.
For example, if we use the following reduction between
$\mathcal{C}_{[a,b]}$ and $\mathcal{C}^{\mathbb{Z}}_{a\mathbb{Z}+b}$, we 
obtain a basis
allowing the construction of a domain for constraints of the form
$(v_i-v_j\in a\cdot[b,c]+d)$:
\begin{center}$\left\{\begin{array}{ll}\relax
[a,b]\circlearrowleft (c\mathbb{Z}+d) \deltaeq
[&\min\{x\in (c\mathbb{Z}+d)\;|\;x\geq a\},
\\ &
\max\{x\in (c\mathbb{Z}+d)\;|\;x\leq b\}\;]; \\\relax
[a,b]\circlearrowright (c\mathbb{Z}+d) \deltaeq &(c\mathbb{Z}+d)\enspace.
\end{array}\right.$\end{center}

\subsubsection{Failure.}
So far, all seems to work well.
However, one can find some bases used in very common abstract domains
that do not respect Hypotheses~\ref{extbasisdef}.
For example, the {\em sign} basis \cite{ai}
$\mathcal{C}_{\pm}=\{\;\bot,\;]-\infty,0],\;[0,0],
\;[0,+\infty[,\;]-\infty,+\infty[\;\}$ and the
open interval basis $\mathcal{C}_{]a,b[}$
do not respect Hypothesis~\ref{extbasisdef}.2.
The {\em interval congruence} basis
$\mathcal{C}_{a\mathbb{Z}+[b,c]}$ (introduced in \cite{intervcong})
does not respect Hypothesis~\ref{extbasisdef}.1
(it is not stable under intersection).
We do not know if it is possible to build weakly relational domains from
such bases.

\subsubsection{Modularity.}
As for non-relational domains, the weakly relational domain family
is simply a generic module functor, taking the very same bases implementation
modules as parameter.

\section{Applications and Future Work}

\subsubsection{Applications.}
So far, this framework has only been implemented as an \textsf{OCaml} 
prototype and tried on a few toy examples.
At program point $(\bullet)$ of the program in Figure~\ref{examplefig}, 
the reduced-product of the zone and zone-congruence domains found the invariant
$(x\leq 5,\;x\equiv 1\;[2])$, which is almost as good as polyhedron and
congruence equality analyses combined (Figure~\ref{resultfig}).
It failed to discover that $(x\geq -5)$; however, the octagon abstract
domain of \cite{mine:ast01} that also uses graph-based algorithms
can do it.

If in the program of Figure~\ref{examplefig} the constant 5 is replaced by a 
variable $m$ the value of which is not
known at analysis time, the analyzer still finds the precise symbolic
invariant $(x\leq m,\;x\equiv m\;[2])$.

\subsubsection{Scalability.}
It is still unknown whether graph-based abstract domains scale up.
Because of the quadratic memory cost, it cannot handle all the
variables of a large program at once; one has to split this set into
{\em packets} in which relational information might be important.
These packets do not need to be disjoint, and one can use pivot
variables to transfer information between packets.
We are currently investigating on such methods.

Because our domain family is relational, it is also adapted to symbolic
and modular analyses.
One can cut down the cost of an analysis and make it incremental by
analyzing separately small pieces of a program \cite{modular}.

\subsubsection{Theoretical Extensions.}
We tried, in this article, to unite some graph-based numerical satisfiability
algorithm and extend them up to an abstract domain, in a united framework.
However, a few graph-based algorithms are not handled here:
the octagon abstract domain \cite{mine:ast01} ($\pm x\pm y\leq c$ constraints)
and Shostak's satisfiability algorithms \cite{shostak}
($\alpha x+\beta y\leq c$).
It would be interesting to unite all those in a general framework and 
derive a numerical abstract domain for constraints of the form
$(\alpha x+\beta y\leq c)$.

\section{Conclusion}

In this paper, we have proposed the systematic construction of a family
of relational domains that represent and manipulate constraints of the form
$(x-y\in C)$.
This construction can be seen as
a {\em functor} lifting non-relational domains to relational ones.
The memory cost of an abstract state is {\em quadratic}, and each transfer
function application performs, at worse, a {\em cubic} number of operations
in the non-relational domain.
The crux of the method is the adaptation of the shortest-path closure
algorithm to a normal form, allowing the derivation of most
abstract operators and transfer functions.

In this framework, we have successfully retrieved the existing DBM domain, 
and constructed new ones.
It is the author's opinion that these domains fill a precision and
complexity gap between former non-relational and relational domains, and
can be used to design medium cost, yet precise, analyses.

\subsubsection{Acknowledgments.}
We would like to thank P.~Cousot, J.~Feret, X.~Rival, and C.~Hymans, as well
as the anonymous referees, for their useful comments.


\bibliographystyle{plain}
\bibliography{bibarticle}


\appendix
\section*{Appendix}

\section{Proof of the Main Theorem}

We present here the complete proof of the main theorem, 
Theorem \ref{closurethm}.
It is the proof of this theorem that motivated the choice of
Hypotheses \ref{extbasisdef}.

Remark that the proof of this theorem is much simpler in the
special case of the interval basis $\mathcal{C}_{[a,b]}$
(see Theorem 2 in the author's master thesis \cite{mine:dea}).

Remark also that part of this theorem for the congruence case
$\mathcal{C}^{\mathbb{Z}}_{a\mathbb{Z}+b}$ is discussed by Toman and Chomicki
in \cite{periodic}, but the proof is somewhat eschewed
(Lemma 2.12). Our proof relies heavily on the fact that 
$\mathcal{C}^{\mathbb{Z}}_{a\mathbb{Z}+b}$ verifies Hypothesis \ref{extbasisdef}.3, which
is not trivial.

\bigskip

{\bf Proof of Theorem \ref{closurethm}.}
\bigskip
\begin{mylist}
\item 
{\bf Claim:} $\Gamma(\vec{m}^{\bigstar})=\Gamma(\vec{m})$.
\finthm

We have 
$\forall k,i,j,\;\vec{m}^{k+1}_{ij}=\vec{m}^k_{ij}\sqcap(\vec{m}^k_{ik}\boxplus\vec{m}^k_{kj})\sqsubseteq\vec{m}^k_{ij}$ 
{\em (Hypothesis \ref{extbasisdef}.1)},
so $\forall k,\;\Gamma(\vec{m}^{k+1})\subseteq\Gamma(\vec{m}^k)$.
Conversely,
$\forall i,j,k,\;(x_0,\ldots,x_{N-1})\in\Gamma(\vec{m}^k)$, we have
$x_k-x_i\in\gamma(\vec{m}^k_{ik})$, and $x_j-x_k\in\gamma(\vec{m}^k_{kj})$.
By summation, 
$x_j-x_i\in\gamma(\vec{m}^k_{ik}\boxplus\vec{m}^k_{kj})$ 
{\em (Hypothesis \ref{extbasisdef}.1)}.
Thus $x_j-x_i\in\gamma(\vec{m}^{k+1}_{ij})$,  and
$\forall k,\;\Gamma(\vec{m}^k)\subseteq\Gamma(\vec{m}^{k+1})$
From these two inequalities, we deduce
$\forall k,\;\Gamma(\vec{m}^{k+1})=\Gamma(\vec{m}^k)$, so
$\Gamma(\vec{m}^{\bigstar})=\Gamma(\vec{m})$.
\finproof

\bigskip
\item 
{\bf Claim:} if $\Gamma(\vec{m})\neq\emptyset$, then
$\vec{m}^{\bigstar}$ is coherent
\finthm

{\em Proof.}
Suppose that $\vec{m}$ is coherent.
We first
prove that
$\forall i,j,\;\vec{m}^{\bigstar}_{ij}=\boxminus\:\vec{m}^{\bigstar}_{ji}$.
By recurrence, one would prove that 
$\forall k,i,j,\;\vec{m}^{k+1}_{ij}=\boxminus\:\vec{m}^{k+1}_{ji}$
using the identity
$\boxminus\:(\vec{m}^k_{ij}\sqcap(\vec{m}^k_{ik}\boxplus\vec{m}^k_{kj}))=
(\boxminus\:\vec{m}^k_{ij})\sqcap((\boxminus\vec{m}^k_{ik})\boxplus(\boxminus\vec{m}^k_{kj}))$ 
{\em (Hypothesis \ref{extbasisdef}.5)}.

\medskip

Now, we know that
$\forall i,\;\vec{m}^{\bigstar}_{ii}\sqsubseteq\vec{m}_{ii}$, so
$\forall i,\;\gamma(\vec{m}^{\bigstar}_{ii})\sqsubseteq\gamma(\vec{m}_{ii})=\{0\}$.
If for some $i$, $\gamma(\vec{m}^{\bigstar}_{ii})\sqsubset \{1\}$, then
$\gamma(\vec{m}^{\bigstar}_{ii})=\emptyset$ and, obviously,
$\Gamma(\vec{m}^\bigstar)=\emptyset$.
This contradicts the fact that $\Gamma(\vec{m})\neq\emptyset$ because
of the preceeding point.

\bigskip
\item
{\bf Lemma 1:}
for any fixed $0\leq i,j\leq N-1,\;0\leq k\leq N,\;$ and path 
$\langle i=i_1,\ldots,i_n=j\rangle$ in $\vec{m}$ such that
$i_l< k$ for $1<l<n$, and
$i_s\neq i_t$ for $1<s<t<n$, we have
$\vec{m}^k_{ij}\sqsubseteq\bigboxplus_{l=1}^{n-1}\vec{m}_{\,i_l\,i_{l+1}}$.
\finthm

{\em Corollary.}
Applying this this lemma with $k=N$, we get:
for all simple paths $\langle i=i_1,\ldots,i_n=j\rangle$,
$\vec{m}^{\bigstar}_{ij}\sqsubseteq\bigboxplus_{l=1}^{n-1}\vec{m}_{\,i_l\,i_{l+1}}$.
\finthm

{\em Proof.}
By recurrence.
The property is obvious for $k=0$ as it is equivalent to
$\vec{m}^0_{ij}\sqsubseteq\vec{m}_{ij}$ and we have $\vec{m}^0=\vec{m}$.
Suppose that the property is true for a $k<N$ and let 
$\langle i=i_1,\ldots,i_n=j\rangle$ be a path satisfying the hypotheses of
the lemma for k+1.
If $\forall l\in\{2,\ldots,n-1\},\;i_l<k$, the property is true by 
recurrence hypothesis and because $\vec{m}^{k+1}_{ij}\sqsubseteq
\vec{m}^k_{ij}$.
On the contrary, if there exists a $l$ such that $i_l\geq k$,
we know that it is unique and that $i_l=k$.
By definition of $\vec{m}^{k+1}$,
we have $\vec{m}^{k+1}_{ij}\sqsubseteq\vec{m}^{k}_{ik}\boxplus
\vec{m}^k_{kj}$.
We obtain the expected result by
applying the recurrence hypothesis
to $\langle i=i_1,\ldots,i_l=k \rangle$ in $\vec{m}^k_{ik}$, and to
$\langle k=i_l,\ldots,i_n=j \rangle$ in $\vec{m}^k_{kj}$, and using
the associativity of $\boxplus$.
\finproof

\bigskip
\item
{\bf Lemma 2:}
if, for some $0\leq i,j<N$,\\
$\gamma(\bigsqcap_{1\leq n,\;\langle i=i_1,\ldots,i_n=j\rangle}
\bigboxplus_{l=1}^{n-1}\vec{m}_{\,i_l\,i_{l+1}})=\emptyset$,
then $\Gamma(\vec{m})=\emptyset$.
\finthm

{\em Proof.}
Suppose that $\gamma(\bigsqcap_{1\leq n,\;\langle i=i_1,\ldots,i_n=j\rangle}
\bigboxplus_{l=1}^{n-1}\vec{m}_{\,i_l\,i_{l+1}})=\emptyset$, but
$\Gamma(\vec{m})\neq\emptyset$.
Take some $(x_0,\ldots,x_{N-1})\in\Gamma(\vec{m})$.
For any path $\langle i=i_1,\ldots,i_n=j\rangle$, we have
$\forall l\in\{1,\ldots,n-1\},\;x_{i_{l+1}}-x_{i_l}\in
\gamma(\vec{m}_{\,i_l\,i_{l+1}})$.
By summation
 $x_j-x_i\in\gamma(\bigboxplus_{l=1}^{n-1}\vec{m}_{\,i_l\,i_{l+1}})$.
Thus $x_j-x_i\in\bigcap_{1\leq n,\;\langle i=i_1,\ldots,i_n=j\rangle}\gamma(
\bigboxplus_{l=1}^{n-1}\vec{m}_{\,i_l\,i_{l+1}})=
\gamma(\bigsqcap_{1\leq n,\;\langle i=i_1,\ldots,i_n=j\rangle}
\bigboxplus_{l=1}^{n-1}\vec{m}_{\,i_l\,i_{l+1}})$ 
{\em (Hypothesis \ref{extbasisdef}.1)}, which is not empty.
\finproof

\bigskip
\item
{\bf Lemma 3:}
if $\forall 0\leq i,j<N$,
$\gamma(\bigsqcap_{1\leq n,\;\langle i=i_1,\ldots,i_n=j\rangle}
\bigboxplus_{l=1}^{n-1}\vec{m}_{\,i_l\,i_{l+1}})\neq\emptyset$, then
$\forall 0\leq i,j<N,\;0\leq k\leq N,\;
\bigsqcap_{1\leq n,\;\langle i=i_1,\ldots,i_n=j\rangle}
\bigboxplus_{l=1}^{n-1}\vec{m}_{\,i_l\,i_{l+1}}\sqsubseteq \vec{m}^k_{ij}$.
\finthm

{\em Corollary.}
When we set $k=N$ in the lemma, we get
$\forall i,j,\;
\bigsqcap_{\langle i=i_1,\ldots,i_n=j\rangle}
\bigboxplus_{l=1}^{n-1}\vec{m}_{\,i_l\,i_{l+1}}\sqsubseteq \vec{m}^{\bigstar}_{ij}$.
\finthm

{\em Proof.}
By recurrence.
If $k=0$, then we have $\vec{m}_{ij}\sqsubseteq\vec{m}^0_{ij}$ because
$\vec{m}^0=\vec{m}$, so {\em a fortiori} the lemma is true.
Suppose that the property is true for $k<N$.
To prove the property for $k+1$, we only have to prove that
$\bigsqcap_{\langle i=i_1,\ldots,i_n=j\rangle}
\bigboxplus_{l=1}^{n-1}\vec{m}_{\,i_l\,i_{l+1}}\sqsubseteq 
\left(\vec{m}^k_{ik}\boxplus\vec{m}^k_{kj}\right)$.

By anti-monotonicity of $\subseteq$ in $\mathcal{P}(\mathcal{C})$ 
($A\subseteq B\subseteq\mathcal{C}\;\Longrightarrow\;\bigsqcap A\sqsupseteq\bigsqcap B$),
we only consider the set of paths from $i$ to $j$ that pass through variable
$k$:\\
$\begin{array}{l}
\bigsqcap_{\langle i=i_1,\ldots,i_n=j\rangle}
\bigboxplus_{l=1}^{n-1}\vec{m}_{\,i_l\,i_{l+1}}\\
\quad\sqsubseteq
\bigsqcap_{\langle i=i_1,\ldots,i_m=k,\ldots,i_n=j\rangle}
\left((\bigboxplus_{l=1}^{m-1}\vec{m}_{\,i_l\,i_{l+1}})
\boxplus
(\bigboxplus_{l=m}^{n-1}\vec{m}_{\,i_l\,i_{l+1}})\right)\\
\quad=
\left(\bigsqcap_{\langle i=i_1,\ldots,i_n=k\rangle}
(\bigboxplus_{l=1}^{n-1}\vec{m}_{\,i_l\,i_{l+1}})\right)
\boxplus\\
\quad\quad
\left(\bigsqcap_{\langle k=i_1,\ldots,\ldots,i_m=j\rangle}
(\bigboxplus_{l=1}^{m-1}\vec{m}_{\,i_l\,i_{l+1}})\right).
\end{array}$

The last equality comes from Hypothesis \ref{extbasisdef}.4 thanks to
$\forall i,j,$
$\gamma(\bigsqcap_{1\leq n,\;\langle i=i_1,\ldots,i_n=j\rangle}
\bigboxplus_{l=1}^{n-1}\vec{m}_{\,i_l\,i_{l+1}})\neq\emptyset$,

To obtain the result, we apply the recurrence hypothesis to
$\vec{m}^k_{ik}$ and $\vec{m}^k_{kj}$.
\finproof

{\em Remark:} the restricted distributivity of $\sqcap$ over $\boxplus$
is crucial in the proof of this lemma.

\bigskip
\item
{\bf Lemma 4:}
if $\exists i,\;0\notin\gamma(\vec{m}^{\bigstar}_{ii})$, then
$\Gamma(\vec{m})=\emptyset$.
\finthm

{\em Proof.}
Suppose that for some $i$, $0\notin\gamma(\vec{m}^{\bigstar}_{ii})$.
This means that $\forall x_i\in\mathbb{I}$, 
$x_i-x_i\notin\gamma(\vec{m}^{\bigstar}_{ii})$,
so $\Gamma(\vec{m}^{\bigstar})=\emptyset$.
By Theorem \ref{closurethm}.1, we get $\Gamma(\vec{m})=\emptyset$.
\finproof

\item
{\bf Lemma 5:}
if $\forall i,\;0\in\gamma(\vec{m}^{\bigstar}_{ii})$, then
$\forall i,j$,\\ 
$\left(
\bigsqcap_{\langle i=i_1,\ldots,i_n=j\rangle}
\bigboxplus_{l=1}^{n-1}\vec{m}_{\,i_l\,i_{l+1}}
\right)
=
\left(
\bigsqcap_{\begin{array}{l}
\mbox{{\scriptsize$\langle i=i_1,\ldots,i_n=j\rangle$}}\\
\mbox{{\scriptsize simple path}}\end{array}}
\bigboxplus_{l=1}^{n-1}\vec{m}_{\,i_l\,i_{l+1}}
\right).$
\finthm

{\em Proof.}
The $\sqsubseteq$ part of the equality is a direct consequence of
the fact that $\sqcap$ is $\subseteq$-anti-monotonic for elements of 
$\mathcal{P}(\mathcal{C})$.

For the $\sqsupseteq$ part, we prove that,
for each path with at least one cycle in it, there exists
a path with one simple cycle less which has a smaller $\boxplus$ sum.
Let $\langle i=i_1,\ldots,i_s,\ldots,i_t=i_s,\ldots,i_n=j\rangle$ be a
path and $\langle i_s,\ldots,i_t=i_s\rangle$ a simple cycle in it.
By Lemma 1, $\bigboxplus_{l=s}^{t-1}\vec{m}_{\,i_l\,i_{l+1}}\sqsupseteq
\vec{m}^{\bigstar}_{ii}$.
By hypothesis, we have $0\in\gamma(\vec{m}^{\bigstar}_{ii})$.
Thus,
$0\in\gamma(\bigboxplus_{l=s}^{t-1}\vec{m}_{\,i_l\,i_{l+1}})$,
and
$\left(\bigboxplus_{l=1}^{n-1}\vec{m}_{\,i_l\,i_{l+1}}\right)\sqsupseteq
\left(\bigboxplus_{l=1}^{s-1}\vec{m}_{\,i_l\,i_{l+1}}\right)\boxplus
\left(\bigboxplus_{l=t}^{n-1}\vec{m}_{\,i_l\,i_{l+1}}\right)$.
\finproof

\bigskip
\item
{\bf Lemma 6:}
if $\Gamma(\vec{m})\neq\emptyset$, then
$\forall i,j,\;\vec{m}^{\bigstar}_{ij}=
\bigsqcap_{\langle i=i_1,\ldots,i_n=j\rangle}
\bigboxplus_{l=1}^{n-1} \vec{m}_{\,i_l\,i_{l+1}}$,
$\forall i,j,k,\;\vec{m}^{\bigstar}_{ij}\sqsubseteq
\vec{m}^{\bigstar}_{ik}\boxplus\vec{m}^{\bigstar}_{kj}$, and
$\vec{m}^{\bigstar}{}^{\bigstar}=\vec{m}^{\bigstar}$.
\finthm

{\em Proof.}
Suppose that $\Gamma(\vec{m})\neq\emptyset$.
By Lemma 2, 
$\forall i,j$, $\gamma(\bigsqcap_{1\leq n,\;\langle i=i_1,\ldots,i_n=j\rangle}
\bigboxplus_{l=1}^{n-1}\vec{m}_{\,i_l\,i_{l+1}})\neq\emptyset$.
Thus, we can apply Lemma 1 and 3 to get
$\forall i,j,\;
\bigsqcap_{\langle i=i_1,\ldots,i_n=j\rangle}
\bigboxplus_{l=1}^{n-1}\vec{m}_{\,i_l\,i_{l+1}}
\sqsubseteq 
\vec{m}^{\bigstar}_{ij}
\sqsubseteq
\bigsqcap_{\begin{array}{l}
\mbox{{\scriptsize$\langle i=i_1,\ldots,i_n=j\rangle$}}\\
\mbox{{\scriptsize simple path}}\end{array}}
\bigboxplus_{l=1}^{n-1}\vec{m}_{\,i_l\,i_{l+1}}$.

By Lemma 4, $\forall i,\;0\in\gamma(\vec{m}^{\bigstar}_{ii})$.
Thus, we can apply Lemma 5 to get
$\forall i,j$, $\vec{m}^{\bigstar}_{ij}=
\bigsqcap_{\langle i=i_1,\ldots,i_n=j\rangle}
\bigboxplus_{l=1}^{n-1}\vec{m}_{\,i_l\,i_{l+1}}
=
\bigsqcap_{\begin{array}{l}
\mbox{{\scriptsize$\langle i=i_1,\ldots,i_n=j\rangle$}}\\
\mbox{{\scriptsize simple path}}\end{array}}
\bigboxplus_{l=1}^{n-1}\vec{m}_{\,i_l\,i_{l+1}}$.

\medskip

Applying a method similar to the one used in Lemma 3, we get:
$\forall i,j,k$,
$\begin{array}{lll}
\vec{m}^{\bigstar}_{ij}&=&
\bigsqcap_{\langle i=i_1,\ldots,i_n=j\rangle}
\bigboxplus_{l=1}^{n-1}\vec{m}_{\,i_l\,i_{l+1}}\\
&\sqsubseteq&
\bigsqcap_{\langle i=i_1,\ldots,i_m=k,\ldots,i_n=j\rangle}
\bigboxplus_{l=1}^{n-1}\vec{m}_{\,i_l\,i_{l+1}}\\
&=&
\left(\bigsqcap_{\langle i=i_1,\ldots,i_m=k\rangle}
\bigboxplus_{l=1}^{n-1}\vec{m}_{\,i_l\,i_{l+1}}\right)
\boxplus
\left(\bigsqcap_{\langle k=i_1,\ldots,i_n==j\rangle}
\bigboxplus_{l=1}^{n-1}\vec{m}_{\,i_l\,i_{l+1}}\right)\\
&=&\vec{m}^{\bigstar}_{ik}\boxplus\vec{m}^{\bigstar}_{kj}.
\end{array}$

\medskip

Using 
$\forall i,j,k,\;\vec{m}^{\bigstar}_{ij}\sqsubseteq\vec{m}^{\bigstar}_{ik}\boxplus\vec{m}^{\bigstar}_{kj}$
in the definition of $\vec{m}^{\bigstar}{}^{\bigstar}$, we get, by recurrence
$\forall i,j,k\;(\vec{m}^{\bigstar})^{k+1}_{ij}=(\vec{m}^{\bigstar})^{k}_{ij}$.
So, $\vec{m}^{\bigstar}{}^{\bigstar}=\vec{m}^{\bigstar}$.
\finproof

\bigskip
\item
{\bf Lemma 7:}
if $\Gamma(\vec{m})=\emptyset$, then
$\exists i,\;0\notin\gamma(\vec{m}^{\bigstar}_{ii})$.
\finthm

{\em Proof.}
We prove this property by recurrence on the size $N$ of the matrix.

If $N=1$, we have obviously $\Gamma(\vec{m})=\{(0)\}\iff
0\in\gamma(\vec{m}_{00})$, and 
$\Gamma(\vec{m})=\emptyset\iff 0\notin\gamma(\vec{m}_{00})$.
By definition, we have 
$\vec{m}^{\bigstar}_{00}=\vec{m}_{00}\sqcap(\vec{m}_{00}\boxplus\vec{m}_{00})$,
so $0\in\gamma(\vec{m}_{00})\iff 0\in\gamma(\vec{m}^{\bigstar}_{00})$.

Suppose the property is true for some $N$.
Let $\vec{m}$ be a matrix of size $N+1$ such that
$\forall i,\;0\in\gamma(\vec{m}^{\bigstar}_{ii})$, we prove that 
$\Gamma(\vec{m})\neq\emptyset$.
Let $\vec{m}'$ be the matrix of size $N$ constructed as follows:
$\forall i,j<N,\vec{m}'_{ij}=\vec{m}_{(i+1)\:(j+1)}\sqcap
(\vec{m}_{(i+1)\:0}\boxplus\vec{m}_{0\:(j+1)})$.
We have $\forall i,j,\;\vec{m}'_{ij}=\vec{m}^1_{(i+1)\:(j+1)}$, so
$\forall i,j,\;\vec{m}'{}^{\bigstar}_{ij}=\vec{m}^{\bigstar}_{(i+1)\:(j+1)}$.
We deduce that $\forall i,\;0\in\gamma(\vec{m}'{}^{\bigstar}_{ii})$ and, by
recurrence hypothesis, $\Gamma(\vec{m}')\neq\emptyset$.
Let us take $(x_1,\ldots,x_N)\in\Gamma(\vec{m}')$.
$\forall 1\leq i,j,\;x_j-x_i\in\gamma(\vec{m}'_{(i-1)\:(j-1)})
\subseteq\gamma(\vec{m}_{ij})$.

Let us prove that we can choose $x_0$ such that $\forall i$,
$x_0-x_i\in\gamma(\vec{m}_{i0})$, and $x_i-x_0\in\gamma(\vec{m}_{0i})$.
This will prove that 
$(0,x_1-x_0,\ldots,x_N-x_0)\in\Gamma(\vec{m})$, and so
$\Gamma(\vec{m})\neq\emptyset$.

First remark that $x_i-x_0\in\gamma(\vec{m}_{0i})\iff x_0-x_i\in\gamma(\boxminus\vec{m}_{0i})\iff x_0-x_i\in\gamma(\vec{m}_{i0})$.
Consider the set 
$C=\gamma(\bigsqcap_{1\leq i} (\{x_i\}\boxplus\vec{m}_{i0}))$.
Then $C\neq\emptyset$, or else, by Hypothesis \ref{extbasisdef}.3
there exists $i,j\geq 1$ such that
$\gamma((x^\sharp_i\boxplus\vec{m}_{i0})\sqcap(x^\sharp_j\boxplus\vec{m}_{j0}))
=\emptyset$,
that is to say $x_j-x_i\notin 
\gamma(\vec{m}_{i0}\boxplus (\boxminus\vec{m}_{j0}))=
\gamma(\vec{m}_{i0}\boxplus \vec{m}_{0j})$,
which is absurd because
$x_j-x_i\in\gamma(\vec{m}'{}^{\bigstar}_{(i-1)\:(j-1)})\subseteq
\gamma(\vec{m}'_{(i-1)\:(j-1)})
\subseteq\gamma(\vec{m}_{i0}\boxplus\vec{m}_{0j})$.
So $C$ is not empty and we simply choose any $x_0\in C$.
\finproof

{\em Remark:} the fact that we can represent singletons, and the stability of
$\boxplus$ are crucial in the proof of this lemma.

\bigskip
\item
{\bf Claim:}
if $\Gamma(\vec{m})\neq\emptyset$, then
$\forall i_0\neq j_0$ and 
$c\in\gamma(\vec{m}^{\bigstar}_{i_0 j_0})$, there exists
$(x_0,\ldots,x_{N-1})\in\Gamma(\vec{m})$
such that $x_{j_0}-x_{i_0}=c$.
\finthm

{\em Proof.} By recurrence on $N$.

The case $N=1$ is not of interest.

When $N=2$ and $\Gamma(\vec{m})\neq\emptyset$,
$\Gamma(\vec{m})=\Gamma(\vec{m}^{\bigstar})=
\{\;(x_0,x_1)\;|\;x_0=0,\;x_1-x_0\in\vec{m}^{\bigstar}_{01}\;\}$.
We can choose, without loss of generality, $i_0=0$, $j_0=1$, so
$c\in\gamma(\vec{m}^{\bigstar}_{01})$.
Then, the property is obvious.

Suppose the property is true for some $N>1$ and 
let $\vec{m}$ be a matrix of size $N+1$ with non-empty domain.
We suppose also, without loss of generality, that $i_0,j_0>0$
($N+1>2$, so one can easily ensure $i_0,j_0>0$ using a simple
variable permutation).
We construct $\vec{m}'$ of size $N$ as in Lemma 7:
$\forall i,j< N,\vec{m}'_{ij}=\vec{m}_{(i+1)\:(j+1)}\sqcap
(\vec{m}_{(i+1)\:0}\boxplus\vec{m}_{0\:(j+1)})$.
Recall that $\forall i,j,\;\vec{m}'{}^{\bigstar}_{ij}=\vec{m}^{\bigstar}_{(i+1)\:(j+1)}$,
so, in particular, $c\in\gamma(\vec{m}'{}^{\bigstar}_{i_0-1\:j_0-1})$.

Applying the recurrence hypothesis to $\vec{m}'$, there exists
$(x_1,\ldots,x_{N})\in\Gamma(\vec{m}')$ such that 
$x_{j_0}-x_{i_0}\in c$.
Then, we can find $x_0$, as in Lemma 7, such that 
$(0,x_1-x_0,\ldots,x_N-x_0)\in\Gamma(\vec{m})$ which ends the proof.
\finproof

\end{mylist}

\end{document}